\documentclass[11pt]{JHEP3}

\usepackage{epsfig}

\newcommand{\be}{\begin{equation} }
\newcommand{\ee}{\end{equation} }
\newcommand{\ba}{\begin{eqnarray}}
\newcommand{\ea}{\end{eqnarray}}

\def\tr{{\rm tr}}

\def\I_M{{I_{\scriptscriptstyle M\times M}}}

\preprint{KIAS-P03050\\SNUTP03-008}

\title{Dyonic Instanton as Supertube between D4 Branes}

\author{Seok Kim \\
\\
School of Physics, Seoul National University, Seoul 151-747, Korea \\
calaf2@snu.ac.kr}

\author{Kimyeong Lee \\
Korea Institute for Advanced Study, Seoul 130-722,
Korea\\
klee@kias.re.kr}

\abstract{We study dyonic instantons in (4+1) dimensional
Yang-Mills theory. Especially we consider the most general two
instanton solution given by the Jackiw-Nohl-Rebbi ansatz and find
its dyonic version. By exploring the zeros of the Higgs field, we
rederive the porism structure of triangles in this solution and
also find the magnetic monopole string loop. This leads to the
identification of dyonic instanton with the supertube inserted
between D4 branes. }

\begin{document}

\section{Introduction}

The gauge field configurations with self-dual field strength in
four dimensions are called instantons. In 4+1 dimensional
spacetime, instantons appear as solitons in the theory. Classical
instantons have zero modes some of which describe the sizes of
instantons. When the gauge symmetry is spontaneously broken,
instantons collapse unless they carry nonzero electric charge and
become dyonic instantons\cite{tong,zam,etz}. In type IIA string
theory the low energy dynamics of parallel D4 branes lying close
to each other is described by 4+1 dimensional Yang-Mills theory
with 16 supersymmetries, while BPS D0-branes on D4 branes are
realized as BPS instantons in this theory. As fundamental strings
connecting parallel branes are electrically charged particles in
the broken phase, dyonic instantons would be considered as BPS
composite objects made of D0 branes and fundamental strings
connecting D4 branes. Refs.\cite{pz,ly} also treat various aspects
of dyonic instantons or calorons.

On the other hand, there has been some study of supertubes which
are 1/4 BPS objects made of fundamental strings, D0 branes and D2
branes in type IIA string theory\cite{mt}. They are 2-dimensional
tube-like object with arbitrary shape of cross section\cite{mnt}.
It has been shown that supertubes can end on a single D4 brane,
while remaining BPS\cite{kmpw}. Supertubes can also be inserted
between two parallel D4 branes, remaining BPS. A BPS configuration
with infinite number of dyonic instantons and parallel magnetic
monopole and anti-magnetic monopole strings has been
found\cite{bl}, which can be interpreted as a supertube made of
parallel D2 anti-D2 branes. One would expect that general dyonic
instantons should be interpreted as supertubes inserted between D4
branes.

However dyonic instantons of 't Hooft type studied in
Ref.~\cite{tong,etz} do not show magnetic monopole string, or
tube-like structure. These solutions seem to describe collapsed
tubes. In this paper we explore the most general two instanton
solution in the SU(2) gauge theory, which is given by the
Jackiw-Norl-Rebbi ansatz. We find the solution of the covariant
Laplace equation for the adjoint scalar field, which leads to the
BPS dyonic solutions. The Higgs field configuration shows how two
D4 branes are deformed near the dyonic instanton. Especially, the
set of points where the Higgs field vanishes is shown to be a
closed magnetic monopole loop, revealing a tube-like structure
connecting two D4 branes. This allows us to interpret the dyonic
instanton as a supertube interpolating two D4 branes.

Supertube, which is a tube-like 1/4 BPS configuration with a
translation symmetry along the tube, have been studied initially
in the DBI action of D2 branes in type IIA theory\cite{mt}. Later,
they have been studied in the matrix theory\cite{ddbar} and in the
supergravity\cite{sugra}. Various configurations like parallel
cylindrical tubes\cite{bl2} and D2 anti-D2 pair\cite{ddbar} have
been found, while it was also shown that the cross section of a
supertube can take arbitrary shape\cite{mnt}. In energetics the
supertube energy can be assigned to those of fundamental strings
and D0 branes, while the bound energy of D0 and fundamental
strings to D2 branes cancel the energy of D2 branes. Hence D2
brane cross section can take arbitrary shape, maintaining BPS
condition. In addition supertubes can end on D4 branes, reducing
the supersymmetry to 1/8 of the superstring theory, and so
carrying 4 supersymmetries. Supertubes can also be inserted
between D4 branes. The obvious field theoretic interpretation
seems to be dyonic instantons in 4+1 dimensional Yang-Mills theory
on D4 branes. This analysis can be done in the DBI action.

The low energy dynamics of two closely lying D4 branes is
described by SU(2) gauge theory with 16 supersymmetries.
Supertubes inserted between D4 branes are expected to be
represented as dyonic instantons in the corresponding field
theory. The Higgs field describes the profile of the deformation
of D4 branes near dyonic instantons. The shape of supertubes
between D4 branes should appear as the zero points of the Higgs
field, where two D4 branes meet. Previous work on dyonic
instantons based on the 't Hooft ansatz has shown that D4 branes
meet on isolated points, instead of some loop. However the 't
Hooft type solutions are not the most general instanton solution.
The tube-like structure may appear in the more general dyonic
instanton solutions.

In this work, we exlore the most general two instanton solution
given by the Jackiw-Nohl-Rebbi (JNR) ansatz\cite{jnr}. The JNR
ansatz for two instantons are characterized by three scale
parameters and three positions in four dimensional space. These
threes-some quantities are related to each other by the local
gauge transformation. Especially the triangle connecting three
positions defines a circle on which three points lie and an
ellipse which three sides of the triangle touches
tangentially\cite{am}. Under a local gauge transformation, the
circle and the ellipse remain invariant, while the triangle
changes on circle. This local gauge transformation is
characterized by a single parameter and this one parameter family
of triangle is called a porism of triangles. As the magnetic
monopole string we find here is not the ellipse which appears in
the porism, the physical meaning of the ellipse remains obscure.

Our method for analysis is the ADHM construction of
instantons\cite{adhm}. The ADHM method for the 't Hooft solutions
has been known. We find its generalization for the JNR solutions,
which is new as far as we know. Then we use the ADHM method to
solve the covariant Laplace equation for the adjoint scalar field.
The ADHM version of the Laplace equation is a matrix equation
which can be solved easily in the two instanton case\cite{dhkm}.

In Sec.2, we summarize the known physics of dyonic instantons in
the spontaneously broken $SU(2)$ gauge theory. We argue that
magnetic monopole string appears naturally along the zeros of the
Higgs field. In Sec.3, we summarize the ADHM method and introduce
the ADHM data for the JNR instanton solutions. We set up the
matrix equation for the solution of the covariant Laplace
equation. In Sec.4, we find the Higgs field solution explicitly.
We rederive the porism of the parameters and find magnetic
monopole strings. Also we investigate the 't Hooft solution limit
to see explicitly that the collapsed tube interpretation of 't
Hooft solution is correct. We conclude with some remarks in Sec.5.

\section{'t Hooft and Jackiw-Nohl-Rebbi Instantons}

We start with the 5-dimensional 16 supersymmetric Yang-Mills
theory with $SU(N)$ gauge group, which describes the low energy
dynamics of $N$ parallel D4 branes whose distances between them
are smaller than the string scale. The Lagrangian is given in the
standard form. The supersymmetry implies a Bogomolnyi bound on the
energy functional. With only a single scalar field being excited,
the bosonic part of the energy functional can be reexpressed as
\begin{equation}
{\mathcal E}=\frac{1}{e^{2}}\int d^{4}x\ {\rm
tr}\left[\frac{1}{4}(F_{\mu\nu}- \tilde{F}_{\mu\nu})^{2}+(E_{\mu}-
D_{\mu}\phi)^{2}
+(D_{0}\phi)^{2}\right]+\frac{8\pi^{2}}{e^{2}}\kappa
+\frac{2}{e^2}Q_{e},
\end{equation}
where $\mu,\nu=1,2,3,4$ and $E_{\mu}=F_{\mu 0}$. The quantities
\be \kappa = \frac{1}{16\pi^2}\int d^4x\; {\tr} F_{\mu\nu}
\tilde{F}_{\mu\nu}, \;\; Q_e= \int d^4x \;
\partial_\mu {\tr } E_\mu \phi ,
\ee
are the instanton number and the energy due to the electric
charge, respectively. The above expression leads to the energy
bound when $\kappa \ge 0$ and $Q_e\ge 0$. With the change of signs
in the energy expression one can give the bound on the energy for
other cases. The energy bound is saturated by the configuration
satisfying
\begin{equation}\label{BPS}
F_{\mu\nu}=\tilde{F}_{\mu\nu}\ ,\ \ E_{\mu}= D_{\mu}\phi\ ,\ \
D_{0}\phi=0.
\end{equation}
The first equation is the self-dual equation for the field
strength, whose solutions are instantons. It is characterized by
the instanton number $\kappa$. When the gauge group is $SU(2)$, it
is known that a general $\kappa$ instanton solution has $8\kappa$
zero modes modulo local gauge transformations. Thus a $\kappa$
instanton solution is characterized by $8\kappa$ moduli
parameters, among which three will be the global gauge
transformation for the gauge group $SU(2)$.

The field configurations should satisfy the Gauss law constraint
\be \label{gauss} D_\mu E_\mu + i[\phi, D_0\phi]= 0. \ee
The last two equations of Eq.~(\ref{BPS}) with the Gauss law
constraint can be combined to be
\begin{equation}\label{BPS-reduced}
\ D_{\!\mu}\!D_{\!\mu}\phi=0.
\end{equation}
(For the BPS configurations satisfying Eqs. (\ref{BPS}) and
(\ref{gauss}), we can choose the gauge where the fields are
time-independent. In this gauge $A_0=\phi$ for the BPS case.) The
above equation is the covariant Laplace equation for adjoint
scalar field in the background of the instanton solution. For this
configuration, the electric charge contribution to the energy
becomes
\be Q_e = \int_{S^3_{\infty}} dS^\mu \;\tr \left( \phi
\partial_\mu \phi\right) =\lim_{x\rightarrow \infty} 2\pi^2 x^2 x_\mu
\tr (\phi \partial_\mu \phi). \ee

The simplest type of self-dual solutions is obtained with the 't
Hooft ansatz
\begin{equation}\label{thooft}
A_{\mu}(x)= \frac{i}{2} \sigma^a \bar{\eta}_{\mu\nu}^{a}
\partial_{\nu}\log H(x) ,\ \ \
\end{equation}
where $\bar{\eta}_{\mu\nu}^{a}$ is the anti-self-dual 't Hooft
tensor. The above ansatz leads to the self-dual field strength if
the field $H(x)$ satisfies $(\partial_\mu^2 H)/H = 0 $. For the 't
Hooft-type solutions
\be \label{Higgs0} H(x)\equiv
1+\sum_{i=1}^{\kappa}\frac{s_i}{|x-a_i|^{2}} . \ee
This solution describes $\kappa$ instantons with $5\kappa $
parameters besides three global gauge parameters.

In the background of the 't Hooft instanton solution, the
covariant Laplace equation has the solution
\begin{equation}\label{Higgs1}
\phi(x)=\frac{q}{H(x)}\ , \ \ \
\end{equation}
which has the asymptotic behaviour $\phi(x)\rightarrow q$ as $
|x|\rightarrow\infty$\cite{tong,etz}. The electric charge energy
for this configuration is
\be Q_e = 2\pi^2 s_\Sigma \; \tr q^2 , \ee
where $s_\Sigma = \sum_{i=1}^\kappa s_i $. This solution $\phi$ is
intrinsically abelian and has zeros at $\kappa$ points
$x=a_1,\cdots, a_k$. The $\phi$ field describes the deformation of
$D4$ branes along a transverse direction. Asymptotically two D4
branes are separated from each other by $\sqrt{\tr{q^2}}$ in the
string scale. They come together at isolated points. There is no
indication of supertubes of finite size connecting two D4 branes.
The 't Hooft solutions seem to describe collapsed supertubes
connecting D4 branes.

Apparently there exists another type of the solutions for the
above ansatz (\ref{thooft}), the so-called Jackiw-Nohl-Rebbi
solution with
\begin{equation}\label{jnr}
H(x)\equiv \sum_{i=0}^{\kappa}\frac{s_i}{|x-a_i|^{2}}.
\end{equation}
This solution describes a configuration of $\kappa$ instantons.
After taking out the overall constant scale of $H$, the number of
parameters of this solution seems naively $5\kappa+4$, besides the
three global gauge transformation parameters. When the points
$a_i$ lie on a single circle or line, an additional constraint due
to a local gauge transformation appears, making one less
parameters to be independent. For the $\kappa=2$ case, the
solution is characterized by three positions $a_i$ in the space
and they always lie on a circle. Thus, the JNR solution with
$\kappa=2$ has 13 instead of 14 independent parameters. With the
additional three parameters for the global gauge parameters, this
JNR solution has 16 parameters, which is exactly the number of
zero modes for the general two instantons.

When we consider the dyonic version, the solution (\ref{Higgs1})
for the covariant scalar equation does not have good asymptotic
value as it diverges. Thus, one needs to find more general
solution. The key task in this work is to find and explore the
solution with the right boundary condition.

There are several aspects of dyonic instantons which are generic.
As the gauge symmetry is spontaneously broken from SU(2) to U(1),
one can generalize the 't Hooft $U(1)$ field strength to 4+1
dimension:
\be G_{MN} = \hat{\phi}^a F^a_{MN} -\epsilon^{abc} \hat{\phi}^a
(D_M \hat{\phi})^b (D_N \hat{\phi})^c .\ee
This becomes $G_{\mu\nu} = \partial_\mu A^3_\nu -\partial_\nu
A^3_\mu$ for a constant $\hat{\phi}^a=\delta^{a3}$. One can define
the $U(1)$ electric charge current $j^\mu$ by
\be j^M = \partial_N G^{NM}, \ee
which is conserved manifestly. It is known that one can define the
magnetic string current $m_{MN}$ as
\be m^{MN} \equiv \frac{1}{2}\epsilon^{MNPQR}
\partial_P G_{QR} , \ee
by generalizing the magnetic monopole current in 3+1 dimensions.
It is manifestly conserved. On the other hand one can define the
covariant topological string current
\be k^{MN} = - \frac{1}{8\pi}\epsilon^{MNPQR} \epsilon^{abc} D_P
\hat{\phi}^a D_Q \hat{\phi}^b D_R \hat{\phi}^c, \ee
which is related to the magnetic monopole current by the relation
$k^{MN}=m^{MN}/4\pi$. The topological current vanishes everywhere
except at the zeros of $\phi$. Unless the zeros of the Higgs field
is highly degenerated, the topological current would not vanish.
As we will see, the zeros of the Higgs field in instanton
background is generically a closed loop and so can be interpreted
naturally as a magnetic monopole string.

Another interesting quantity is the angular momentum which is
\be L_{\mu\nu} = \int d^4 x ( x_\mu T^0_{\; \nu} - x_\nu T^0_{\;
\mu} ), \ee
where $T^\mu_{\;\nu}$ is the conserved energy momentum tensor. For
't Hooft type self-dual dyonic instantons the angular momentum
simplifies to be anti-selfdual. There is no indication of such a
simplication of the angular momentum for more general dyonic
instanton configurations we study here.

\section{The ADHM Method}

To obtain the scalar solution in the JNR background, we employ the
ADHM method. As we will see, the answer is quite nontrivial and
could not be reached easily by an ansatz. Let us start with a
brief review of the ADHM method. For the $\kappa$-instanton
configurations in the SU(2) gauge theory, we start with $\kappa$
dimensional row vectors $\Lambda_\mu$ and $\kappa\times \kappa$
Hermitian matrices $\Omega_\mu$, and introduce a quaternionic
$(\kappa+1)\times \kappa$ matrix
\begin{equation}\label{standard-data}
\Delta(x)=\left(\begin{array}{c}\Lambda_\mu e_\mu \\ \
(\Omega_{\mu} -x_\mu)e_{\mu}
\end{array}\right),
\end{equation}
where $e_{\mu}\equiv( i\vec{\sigma}, \ 1)$. The $x$ dependence of
$\Delta(x)$ is shown explicitly in the above equation. Note that
$e_\mu\bar{e}_\nu = \delta_{\mu\nu} +i\eta^{a}_{\mu\nu}\sigma^a$
with the selfdual 'tHooft tensor $\eta^a_{\mu\nu}$. We require
that $\Delta^\dagger \Delta$ is proportional to the identity
element of quaternion and invertible. This leads to three
quadratic matrix equations on $\Lambda_\mu$ and $\Omega_\mu$, and
we call those satisfying this condition to be the ADHM data. Then
we find a $\kappa+1$ dimensional quaternionic column vector $v(x)$
such that $\Delta^\dagger(x) v(x)=0$ and $v^\dagger v = 1$, which
leads to the selfdual instanton configuration $A_\mu = v^\dagger
\partial_\mu v$. There is a $U(\kappa)$ group action on the ADHM
data, whose parameters do not appear in the instantons. When one
counts the number of free parameters which determine the instanton
configuration, one finds that $8\kappa$ is the number of zero
modes for $\kappa$ instanton configuration in the SU(2) gauge
group case.

To find the solution of the covariant Laplace equation in the
adjoint representation, we start with an ansatz as\cite{dhkm,tong}
\begin{equation}\label{higgs-sol}
\phi=v^{\dag}\left(\begin{array}{cc}
q & 0\\
0& p \end{array}\right)v,
\end{equation}
where $q$ is the asymptotic Higgs value and also an element of the
$SU(2)$ algebra. The $\kappa\times \kappa$ matrix $p$ belongs the
identity element of quaternion and satisfies the constraint,
\begin{equation}\label{zero-constraint}
-[\Omega_\mu,[\Omega_\mu,p]] - \left\{ \Lambda^\dagger_\mu
\Lambda_\mu,p\right\} -2i\eta^a_{\mu\nu} q^a \Lambda^\dagger_\mu
\Lambda_\nu = 0 ,
\end{equation}
where $q=q^a\sigma^a$. Here we choose the $\phi$ to be a hermitian
matrix and so are $q$ and $p$.

The $\kappa=2$ ADHM data has been constructed a long
ago\cite{cws}. But it does not lead to the JNR ansatz. Here we
introduce the ADHM data for the JNR instanton solutions for
$\kappa$ instantons, because geometric interpretation of the data
would be easier. This is achieved by mimicking the ADHM
construction of the 't Hooft solution. With the quaternion
position operators $y_i=(x-a_i)_\mu e_\mu$, the relavant ADHM data
is
\begin{equation}\label{jnr-adhm}
\Delta(x)=\left(\begin{array}{c} y_0\Lambda \\\
-Y_{\kappa\times\kappa}
\end{array}\right) ,
\end{equation}
where $\Lambda$ is a row vector with real number components
$(\lambda_1/\lambda_0, \cdots, \lambda_\kappa/\lambda_0)$ and $Y$
is a quaternionic matrix $Y={\rm diag}(y_1,\cdots, y_k)$. (Taking
the limit $a_{04}=\lambda_{0}\rightarrow\infty$, we recover the
ADHM data for the 't Hooft solution.) This indeed satisfies the
ADHM constraint equations, and the $(\kappa+1)$ column vector
$v(x)$ is chosen such that the $i$-th entry is
\begin{equation}\label{fibre}
v_i = \frac{\lambda_i y_i}{y_i^2} \frac{1}{\sqrt{H}} ,
\end{equation}
where $y_i^2=|y_i|^2 = \sum_\mu y_{i\mu} y_{i\mu}$. The
corresponding gauge field $A_\mu = \bar{v}\partial_\mu v$ is the
JNR solution with $s_i=\lambda_i^2$. Note that the `permutation
symmetry' between $(y_0,s_0)$, $(y_1,s_1)$, and $(y_2,s_2)$ is
obscure in the ADHM data (\ref{jnr-adhm}) but manifest in the
gauge field.

To get contact with the standard form (\ref{standard-data}), we
need to perform a transformation $\Delta\rightarrow U\Delta K$
with our JNR data (\ref{jnr-adhm}), which carries same information
provided $U\in U(2\kappa+2)$ and $K\in GL(\kappa,C)$. To change
the matrix $\Delta(x)$ into the standard form, we choose $K$ and
$U$ as
\begin{equation}\label{into-standard}
K^{-2} = 1+ \Lambda^T\Lambda,
\;\; U=\left(\begin{array}{cc}u& u\Lambda\\
-K\Lambda^T & K\end{array}\right) ,
\end{equation}
where $u^{-2} = 1+\Lambda\Lambda^T $. Defining the relative
instanton position matrix ${\cal A} ={\rm
diag}(a_1-a_0,\cdots,a_\kappa-a_0)$, the result is
\begin{equation}\label{jnr-adhm2}
U \Delta(x)K=\left(\begin{array}{c}
u\Lambda{\cal A}K\vspace{0.2cm}\\
K{\cal A }K +a_0-x
\end{array}\right).
\end{equation}
Note that the above expression is given in the standard form
(\ref{standard-data}) As we make above transformation on $\Delta$,
$v(x)$ in Eq.(\ref{fibre}) also changes into $Uv(x)$.

It is difficult to solve the constraint (\ref{zero-constraint})
for $p$ with ADHM data (\ref{jnr-adhm2}). If the solution $p$ is
known, the Higgs solution (\ref{higgs-sol}) becomes
\be\label{higgs-sol-exp} \phi(x)=\frac{1}{s_\Sigma
H(x)}\left[\bar{Z}qZ + Q(x)\right] , \ee
where $ s_\Sigma=\sum_{i=0}^\kappa s_i $,
\be\label{force} Z(x)=Z_\mu e_\mu = \sum_{i=0}^\kappa \frac{s_i
y_{i\mu} e_\mu}{y_i^2}, \ee
and
\be \label{qx} Q(x) =
\lambda_{0}^2\Lambda^{T}\left\{\frac{1}{Y}-\frac{1}{y_0}\right\}
K\, p \, K\left\{\frac{1}{\bar{Y}}-
\frac{1}{\bar{y}_0}\right\}\Lambda. \ee
The asymptotic value is $\phi = \bar{U}_0(x)qU_0(x)$ with $U_0(x)
= e_\mu x_\mu/|x|$. The permutation symmetry in $Q(x)$ would be
manifest once we solve the equation (\ref{zero-constraint}) for
$p$.

Before embarking upon the two instanton case, let us check whether
the above solution for the scalar field works for a single
instanton. The JNR solution for a single instanton should be gauge
equivalent to that of 't Hooft type. Equally the scalar field in
the above case should be gauge equivalent to the known solution of
the 't Hooft type. We first note that for a single instanton with
$\kappa=1$,
\be H=\frac{s_0}{y_0^2}+ \frac{s_1}{y_1^2} =
\frac{(s_0+s_1)}{y_0^2 y_1^2} (x-a_x)^2 H_c , \ee
where
\be H_c= 1+ \frac{s_c}{(x-a_c)^2} , \ee
and $s_c= s_0s_1|a_1-a_0|^2/(s_0+s_1)^2$, $a_c =
(s_1a_0+s_0a_1)/(s_0+s_1)$. Also $p$ in
Eq.~(\ref{zero-constraint}) vanishes identically for $\kappa=1$.
We define a global gauge transformation
\be V= \frac{ (a_1-a_0)_\mu e_\mu}{|a_1-a_0|}, \ee
and a local gauge transformation
\be U(x) = \frac{Z_\mu e_\mu}{|Z|} \ee
with $Z_\mu = s_0 y_{0\mu}/y_0^2 + s_1 y_{1\mu}/y_1^2$. One can
show that the gauge field and the adjoint scalar field in the JNR
type with $\kappa=1$ can be rewritten as
\ba && A_\mu = \bar{U}V \left(\frac{i}{2}
\bar{\eta}_{\mu\nu}^a\sigma^a
\partial_\nu \ln H_c \right) \bar{V}U + \bar{U}V\partial_\mu
(\bar{V}U), \nonumber \\
&& \phi = \bar{U}V \left( \frac{\bar{V}q V}{H_c(x)} \right)
\bar{V} U . \ea
This is a local gauge transformation of the 't Hooft type single
instanton solution. The instanton position and scale are given as
$a_{c\mu}$ and $s_c$. The asymptotic value of the scalar field is
gauge equivalent to a constant matrix $\bar{V}qV$.

\section{Two Instanton Solutions}

Now we analyse two equations (\ref{zero-constraint}) and
(\ref{higgs-sol-exp}) for $\kappa=2$. Without loss of generality,
we align the instanton points $a_0,a_1, a_2$ on the 3-4 plane,
i.e., $a_{i1}=a_{i2}=0$. Then $\Omega_\mu = K{\cal
A}_{\mu}K+a_{0\mu}$ appearing in Eq.(\ref{zero-constraint}) is
zero unless $\mu=3,4$, which means that the left hand side of
Eq.(\ref{zero-constraint}) is proportional to $\eta_{34}^{\ \
a}=\delta_{a3}$. Therefore, the matrix $p$ is proportional to
$q^3$, and does not depend on $q^1,q^2$. (With the ansatz
(\ref{thooft}), one has implicitly chosen a global gauge such that
this plane is related to third direction in internal space.)

In order to solve Eq.(\ref{zero-constraint}), we write down an
explicit expression for the matrix $K$ defined in
Eq.(\ref{into-standard}):
\begin{equation}
K=\frac{1}{\Lambda^{\!T}\!\Lambda} \left(\begin{array}{cc}
u\lambda_{1}^{2}+\lambda_{2}^{2}&\lambda_{1}\lambda_{2}(u-1)\\
\lambda_{1}\lambda_{2}(u-1)&\lambda_{1}^{2}+u\lambda_{2}^{2}\end{array}\right)
\ ,
\end{equation}
where $ u=|\lambda_0|/(\lambda_0 + \lambda_1^2+\lambda_2^2)^{1/2}
$. Using this expression, we solve the constraint
(\ref{zero-constraint}) in terms of the $2\times 2$ matrix $p$. As
there is only one explicit $i$ in the last term of
Eq.(\ref{zero-constraint}), the hermitan matrix $p$ should be pure
imaginary and antisymmetric. All we need to compute is then the
single coefficient of two-dimensional antisymmetric matrix. After
some algebra and rearrangement, we find the matrix $p$. To express
the answer in compact form, we introduce two cyclic quantities for
two instantons,
\be {\cal P} = \frac{4 \; q^a \eta^a_{\mu\nu} \; (a_{0\mu}a_{1\nu}
+ a_{1\mu}a_{2\nu}+ a_{2\mu}a_{0\nu}) } {(s_0s_1)^{-1} |a_0-a_1|^2
+ (s_1s_2)^{-1}|a_1-a_2|^2+ (s_2s_0)^{-1}|a_2-a_0|^2 } , \ee
and
\be {\cal F}(x) = \sigma^a \bar{\eta}^a_{\mu\nu} \left(
\frac{y_{0\mu}}{|y_0|^2}\frac{y_{1\nu}}{|y_1|^2} +
\frac{y_{1\mu}}{|y_1|^2}\frac{y_{2\nu}}{|y_2|^2} +
\frac{y_{2\mu}}{|y_2|^2}\frac{y_{0\nu}}{|y_0|^2} \right). \ee
The quantity $Q(x)$ of Eq.(\ref{qx}) is given as $Q(x) = {\cal
P}{\cal F}(x)$ which has the permutation symmetry between
$i=0,1,2$ indices. The resulting scalar field is then
\be \phi(x) = \frac{1}{s_\Sigma H(x)} (\bar{Z} q Z + {\cal P}
{\cal F}) \label{sol}, \ee
where $Z(x) = (s_0y_{0\mu}/y_0^2+ s_1
y_{1\mu}/y_1^2+s_2y_{2\mu}/y_2^2)e_\mu$. We have also checked
explicitly that the above solution satisfies the covariant Laplace
equation. The relative sign between two terms in the numerator is
crucial for the supertube configuration to appear. The solution
(\ref{sol}) is our key result.

With this solution, we find the electric charge energy becomes
\be Q_e = \frac{8\pi^2}{s_\Sigma^2} \left\{ (q^a)^2 \sum_i s_i
s_{i+1} (a_i-a_{i+1})^2 -\frac{4(\sum_i q^a\eta^a_{\mu\nu}
a_{i\mu}a_{(i+1)\nu})^2}{ \sum_j (s_js_{j+1})^{-1}
(a_j-a_{j+1})^2} \right\} ,\ee
where $q=q^a\sigma^a$ and the sum is over $0,1,2$ indices with the
identification $a_3=a_0$, $s_3=a_0$. After putting three points
$a_0, a_1, a_2$ on the 3-4 plane and using the Schwartz inequality
and general property of triangles, one can easily show that
\be Q_e \ge \frac{16\pi^2}{s_\Sigma^2} \frac{ (q^a)^2 (\sum_i
(a_i^2)^2) }{ \sum_i (s_is_{i+1})^{-1} (a_i-a_{i+1})^2} . \ee
This shows that $Q_e$ is positive, as it should be.

Let us now consider the zeros of the Higgs field. Without loss of
generality, one can rotate the coordinate and put three instanton
positions $a_i$ on the 3-4 plane. When $q^3=0$, we note that
${\cal P} $ vanishes. In this case, the zeros of $\phi$ are
determined by the zeros of $Z=Z_\mu e_\mu$ as $H$ is positive
definite and $1/H$ does not vanish anywhere. Due to the symmetry,
the zeros of $Z$ should be also on 3-4 plane. Then $Z$ vanishes at
points where the sum of two dimensional `Coulomb forces' (due to
three positive `charges' $s_0, s_1, s_2$ at $a_0, a_1, a_2$)
vanishes. The zero points are critical points of the potential.
Generically there are two such points. Consider two charges close
to each other and one far apart. The equipotential lines
surrounding two close charges merge together, generating one such
zero point. Then this merged equipotential lines will merge with
those from the third charge, generating another zeros. Later we
will write down the algebraic equation for the zero points which
shows that there are two zero points indeed. On the other hand
when $q^3\ne 0$ and $q^1, q^2=0$, ${\cal P}$ does not vanish.
Again in this case, symmetry implies that the zeros of the scalar
field should be confined to the 3-4 plane. We will see they form a
curve on the 3-4 plane. If all components of $q^a$ do not vanish,
the zeros of the Higgs field do not remain on the 3-4 plane and
would become far more complicated, so we will not pursue this case
here.

\subsection{A derivation of the structure of porism}

The JNR two-instanton solution carries 15 parameters (3
four-positions and 3 scales), two more than 13 moduli as we expect
for two-instanton solution. The overall scale is trivial as it
leaves the potential (\ref{thooft}) invariant. Another unphysical
parameter is related to a local gauge transformation. First put
three points $a_{0\mu}, a_{1\mu}, a_{2\mu}$ on a circle on the 3-4
plane. The lines connecting them forms a triangle inside the
circle. It is known that, under certain local gauge
transformation, the three points $a_0,a_1,a_2$ move along the
circle with speeds proportional to $s_0,s_1,s_2$, respectively.
There exists an ellipse inside the circle which touches all sides
of the one-paramter family of triangles tangentially. This ellipse
also remains invariant under the local gauge transformation. This
fact also determines how the scale parameters should transform
under the local gauge transformation. This one parameter family of
triangles is the so-called porism of triangles\cite{am}. However,
the physical meaning of this ellipse remains obscure.

It seems that the already-known derivation of this result is
somewhat involved. Here we describe another way to derive the
porism by using our Higgs solution (\ref{sol}). As the scalar
field transforms homogeneously under local gauge transformations,
the zero points of the scalar field are gauge invariant.
Especially with $a_{i\mu}$ on the 3-4 plane and $q^3=0$, the
scalar field has at most two isolated zero points on the 3-4
plane, which should be invariant under local gauge
transformations. We aim to derive the deformation rule by studying
a single equation $Z(x)=0$.

For our $\kappa=2$ case, $Z(x)=0$ amounts to a quadratic equation
of $x_\mu e_\mu$ on the 4-3 plane. We put three points on a circle
whose radius is $R$ and whose center is at the origin. In complex
notation, we put $a_{j4}+ia_{j3} = Re^{i\theta_j}$ and a point
$z=x_4+ix_3$ on the 3-4 plane. The zeros of $Z(x)$ satisfy a
complex equation
\begin{equation}\label{quad}
z^2 + C_1 Rz + C_2 R^2= 0 ,
\end{equation}
where
\be C_1= -\sum_{j=0}^2 (1-\frac{s_j}{s_\Sigma}) e^{i\theta_j},
\;\; C_2 = e^{i(\theta_0+\theta_1+\theta_2)} \sum_j
\frac{s_j}{s_\Sigma} e^{-i\theta_j} . \ee
As the above equation is quadratic, there can be at most two
isolated zeros. They are determined by two complex numbers $C_1,
C_2$. We are looking for a change of parameters which leave the
circle and the Higgs zeros invariant. That is an arbitrary
variation of the angles $\theta_i$ and $s_i$ which leaves
$s_\Sigma, C_1, C_2$ invariant. As there are five real paramters
in $s_\Sigma, C_1, C_2$ to fix and six parameters $\theta_i$ and
$s_i$ to vary, there is one free parameter among $\theta_i$ and
$s_i$, which is the porism of triangles.

Under infinitesimal change $\delta \lambda_i$ and $\delta s_i$
such that $\delta s_\Sigma=0$, we note that
\be \delta C_1 - e^{i\sum_i \theta_i} \delta \bar{C}_0 = i
\sum_{j,k} e^{i\theta_j} (-s_j \delta \theta_k + s_j \delta
\theta_k). \ee
The vanising condition of the above term for generic $s_j,
\theta_j $ implies
\be \delta \theta_j = s_j \delta t \ee
with an infinitesimal parameter $t$. As
\be \delta C_1 = \sum_j e^{i\theta_j} \left( \frac{\delta
s_j}{s_\Sigma} -i (1-\frac{s_j}{s_\Sigma} ) \delta \theta_j
\right), \ee
the equation $\delta C_1=0$ implies a complex equation
\be \sum_j e^{i\theta_j } \left( \frac{\delta s_j}{s_\Sigma} -i (
1-\frac{s_j}{s_\Sigma} )s_j \delta t \right) = 0 \ee
for two independent $\delta s_j$. Therefore, the structure of
porism is completely derived from gauge-invariance of the zero
points of the Higgs field.

\subsection{Supertube Connecting D4 branes}

Let us now consider the zeros of the scalar field when $q^3\ne 0$
and $q^1,q^2=0$ with $a_i$ on the 3-4 plane. The zeros of the
scalar field would lie on the 3-4 plane due to the symmetry and is
given by the solutions of a real equation
\be X= \sum_{i=0}^2 (s_i^2 \; y_{i+1}^2 y_{i+2}^2 +2s_i s_{i+1}\;
y_i \cdot y_{i+1}\; y_{i+2}^2 - {\cal P }' y_{i+2}^2 \; y_i \times
y_{i+1} ) = 0 \ee
for a two dimensional vector $x=(x_3,x_4)$, where $y_i=x - a_i$
become two dimensional vectors on the 3-4 plane and
\be {\cal P}' = \frac{4 \; (a_{0\mu}\times a_{1\nu} +
a_{1\mu}\times a_{2\nu}+ a_{2\mu}\times a_{0\nu}) } {(s_0s_1)^{-1}
|a_0-a_1|^2 + (s_1s_2)^{-1}|a_1-a_2|^2+ (s_2s_0)^{-1}|a_2-a_0|^2
}. \ee
Here the cross product is the two dimensional skew product,
$a\times b = a_3 b_4-a_4 b_3$.

\EPSFIGURE[!t]{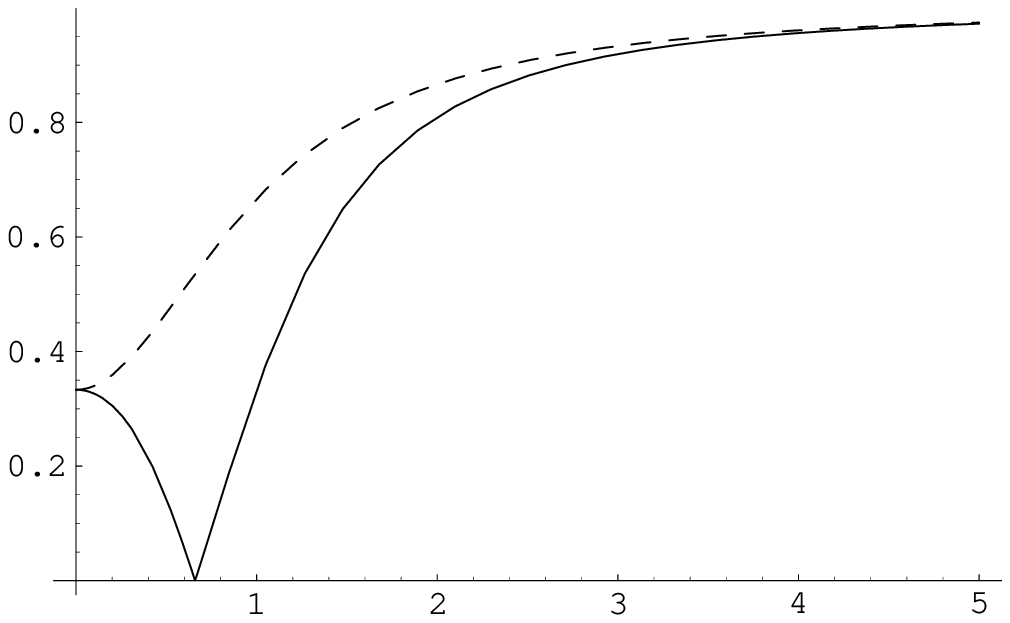}{The Higgs field profile in the unit of
$q^3\sigma^3$: the solid line is along $x^3$ axis and the dashed
line is along $x^1$ axis. }

Let us consider a simple case where three points $a_i$ are the
vertices of an equitriangle, $a_0= (-1,0)R$, $a_1= (1/2,
-\sqrt{3}/2)R$, and $a_2= (1/2, \sqrt{3}/2)R$ on the 3-4 plane,
and all scales are such that $s_i = 1$. Then
\be X= 9 (x^2)^2 + (6 -\frac{3\sqrt{3}}{2} {\cal P}')R^2x^2
-\frac{3\sqrt{3}}{2} {\cal P}' R^4 = 0 .\ee
As ${\cal P}' = 2/\sqrt{3}$ in this case, the above equation has
zero at points
\be x^2 = \frac{\sqrt{13}-1}{6} R^2\ee
which is a square of the radius of a circle. Since the scalar
field has nontrivial expectation value at spatial infinity and has
zeros on a circle, two D4 branes separated along the transverse
direction come together on a circle. There is no special
degeneracy of zeros and so the circle becomes the magnetic
monopole string. We interprete this dyonic configuration as the
supertube inserted between two separated D4 branes. The value of
the Higgs field at the center of the circle or the orgin is
$q^3\sigma^3/3$, which is smaller than the asymptotic value,
regardless the size of the equitriangle. As the size $R$
increases, the electric charge increases and is able to deform the
Higgs field so that the value at the center remains smaller. The
value of the Higgs field along $x^1$ axis and $x^3$ axis with
$R=1$ are drawn in Fig. 1.

Let us now consider a little more general case. Instead we choose
$a_0= (-s_0, 0)$, $a_1= (1/2, -\sqrt{3}/2)$, and $a_2= (1/2,
\sqrt{3}/2)$ on the 3-4 plane with $R=1$. We choose $s_0$ arbitrary
and $s_1=s_2=1$. Then we change the scale $s_0$ from 1 to $\infty$
gradually, which interpolates between the JNR and the 't Hooft
solutions of two instantons. (When we change $s_0$ from 1 to 0, the
solutions interpolate between the JNR and the 't Hooft solution of a
single instanton. The zeros of the Higgs changes acoordingly.)  In the
't Hooft case, the zeros are located at two points $a_1, a_2$. In
Fig. 2, we draw the zeros of the Higgs field for the various values of
$s_0$. For the $s_0=1$, the zeros make a circle as we described
before. The figure shows the gradual deformation of the circle, or the
magnetic monopole string, into two points. (The curve of the zeros is
not convex and is not clearly related to the ellipse of the porism.)
This illustrates that the 't Hooft dyonic instanton indeed represents
collapsed tubes.


\begin{figure}[!t]
\hskip 4cm
\includegraphics[width=0.4\textwidth]{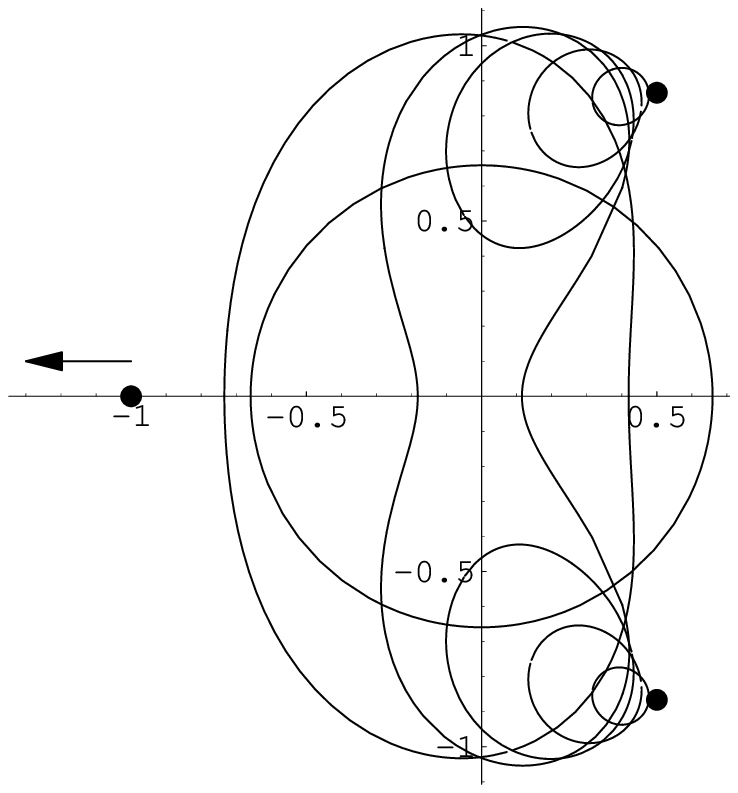}
\caption{Deformation of Magnetic monopole strings for various
values of $s_0$.}
\end{figure}

\section{Conclusion}

We investigated dyonic two instanton configurations with the JNR
ansatz, which is most general two instanton solution. We have
shown that the monopole string structure appears with this
solution, which leads naturally to the interpretation of dyonic
instantons as supertubes between D4 branes. We have exlored the
detail of the porism of triangle in the JNR ansatz and have also
seen how the monopole string deforms as the moduli parameters
change, especially as the JNR solution deforms to 'tHooft solution
of two instantons.

It would be natural to look for how our construction generalizes
to higher instanton number case. At least with the JNR ansatz, it
seems to be doable to find the explicit form of the Higgs field
with some effort. Our preliminary investigation shows that there
is a hierarchy of clustering of terms in the Higgs solution.

By the way, our solution (\ref{sol}) might be useful in a
different direction. To find the moduli space of instantons, one
has to find infinitesimal variation of the gauge field which
satisfies the covariant gauge or the Gauss law. As far as we know,
no explicit moduli space metric has been found for two instanton
solutions. The adjoint Higgs field we found is intimately related
to the global gauge zero modes of instantons. The electric energy
would be the kinetic energy for the global gauge transformation
with the gauge parameters given by $q^a$. It would be a challenge
to find the whole moduli space metric for two instantons.

\acknowledgments{The authors wish to thank Dongsu Bak and Piljin
Yi for the enlightening discussions. S.K is supported in part by
the BK21 project of the Ministry of Education, Korea, and the
Korea Research Foundation Grant 2001-015-DP0085. K.L. is supported
in part by KOSEF 1998 Interdisciplinary Research Grant.}

\end{document}